\documentclass[prb,preprint]{revtex4-1}
\usepackage[active]{srcltx}
\usepackage{graphicx}
\usepackage{bm}
\usepackage{textcomp}
\usepackage{units}
\usepackage{amsmath}

\begin{document}

\def\fr#1#2{\frac{\textstyle #1}{\textstyle #2}}
\def\rd{{\rm d}}
\def\re{{\rm e}}
\def\ri{{\rm i}}
\def\iO{{\it \Omega}}
\title{Dimensionless Units in the SI}

\author{Peter J. Mohr}
\email{mohr@nist.gov}

\author{William D. Phillips}
\email{wphillips@nist.gov}

\affiliation{National Institute of Standards and Technology,
Gaithersburg, MD 20899, USA}

\begin{abstract} 

The International System of Units (SI) is supposed to be coherent.  That
is, when a combination of units is replaced by an equivalent unit, there
is no additional numerical factor.  Here we consider dimensionless units
as defined in the SI, {\it e.g.} angular units like radians or
steradians and counting units like radioactive decays or molecules.  We
show that an incoherence may arise when different units of this type are
replaced by a single dimensionless unit, the unit ``one'', and suggest
how to properly include such units into the SI in order to remove the
incoherence.  In particular, we argue that the radian is the appropriate
coherent unit for angles and that hertz is not a coherent unit in the
SI.  We also discuss how including angular and counting units affects
the fundamental constants.

\end{abstract}

\centerline{\today}

\maketitle

\subsection*{\it A parable of dimensionless units}
{\it

Bert has a turntable operating at a rotation frequency of one radian per
second.   His friend, Ernie, asks Bert, ``At what frequency is your
turntable rotating?''  If Bert were to answer ``0.16 Hz'' Ernie would
know the rotation frequency.  Similarly, if Bert were to answer ``1
radian per second,'' Ernie would know the rotation frequency.  And, if
Bert were to say ``57 degrees per second,'' Ernie would be well
informed.  On the other hand, if Bert were to respond ``The rotation
frequency is 1,''  we would all agree that Ernie would not know the
rotation frequency.  Nor, would a response of  ``1 per second'' be
useful to Ernie, any more than a response of ``57 per second.''   In
order to convey useful information, Bert must tell Ernie the units in
which he is reporting the rotation frequency, including the so-called
“dimensionless” units of cycles or radians or degrees.  The current
formulation of the SI specifically allows the units ``radian'' or
``cycles'' to be replaced by the dimensionless unit ``one,'' and it
allows both radians/second and cycles/second (Hz) to be replaced with
inverse seconds.  Clearly, if Bert had followed this prescription,
allowed by the current SI, he would have left Ernie in the dark about
the rotation frequency.  

If Bert had given an uninformative response about the rotation rate,
Ernie might have asked ``what is the concentration of oxygen in the air
you're breathing?'' Bert could respond ``about $10^{25}$ atoms per cubic
meter'' or ``\,$5\times10^{24}$ molecules per cubic meter.''  These are
clear answers.  A factor-of-two ambiguity would arise if he had not
specified the entity being counted; in fact, the current SI says that
``atoms'' or ``molecules'' are dimensionless units that should be set
equal to ``one''.  If Bert had said ``$5\times10^{24}$ per cubic
meter,'' Ernie might have interpreted that as being the atomic density
and wonder if oxygen deprivation had compromised his friend's mental
acuity.

Such situations allowed by ambiguous units are untenable, which is one
of the main points of this paper.   Fixing the problem is not going to
be easy, as evidenced by the fact that it has persisted for so many
years after the institution of the SI in 1960.  This situation has led
to such problematic pronouncements as ``The radian and steradian are
special names for the number one \ldots''\,\cite{2006SI}. One starting
point is to recognize that replacing “radians” or “cycles” or similar
dimensionless units by ``one'' leads to trouble, and replacing molecules
by ``one'' in expressions for molecular concentration may also lead to
trouble.  These and similar arguments are made explicit below, as are
our suggestions for a revision to the SI that goes a long way toward
solving the problem.}\footnote{The hypothetical conversation in this
section is not meant to suggest that there was an actual such
conversation between Albert Einstein and Ernest Rutherford.}

\section{Introduction}

The International System of Units (SI) defines units that are used to
express the values of physical quantities.\cite{2006SI}  In the
foreseeable future, it is expected that there will be a redefinition of
the SI based on specified values of certain fundamental
constants.\cite{2014050}  This constitutes a dramatic change with one of
the consequences being that there will no longer be a clear distinction
between base units and derived units.\cite{2011191,2008027}  In view of
this change, it is timely to reexamine units in the SI and their
definitions.  One goal is to insure that all such units are coherent,
{\it i.e.,} they comprise a coherent system of units.

In the current SI, various quantities are designated as being
dimensionless.  That is, they are deemed to have no unit or have what
has been called the coherent derived unit ``one''.  In some cases this
designation leads to ambiguous results for these quantities.  In this
paper, we examine units in the SI that are considered dimensionless and
other units not presently included in the SI that might be added to
bring it into closer alignment with widespread scientific usage.

\section{Units and dimensional analysis}

In general, units are used to convey information about the results of
measurements or theoretical calculations.  To communicate a measurement
of a length, for example, the result is expressed as a number and a
unit, which in the SI is
the meter.  The number tells the length in meters of the result of the
measurement.

For simple algebraic calculations involving units, one can write out the
expression and separately collect the units, which may be replaced by an
equivalent unit for convenience.  For example, the kinetic energy $E$ of
a mass $m = 2$ kg moving at a velocity $v = 3$ m/s is calculated as
\begin{eqnarray} 
E = \frac{1}{2}\,mv^2 = \frac{1}{2}\left(2 \mbox{
kg}\right) \left(3 \mbox{ m/s}\right)^2 = \frac{2\cdot3^2}{2} \mbox{
kg}\mbox{ m}^2\mbox{ s}^{-2} =9 \mbox{ J}\,, \label{eq:simex}
\end{eqnarray} 
where J is the symbol for joule, the SI unit of energy.
This calculation illustrates the important principle of the SI that the
units are coherent.  That is, when a combination of units is replaced by
an equivalent unit, there is no additional numerical factor.  For
Eq.~(\ref{eq:simex}), this corresponds to the relation \begin{eqnarray}
\mbox{ kg}\mbox{ m}^2\mbox{ s}^{-2} = \mbox{ J} \, .  \label{eq:enequiv}
\end{eqnarray}

The notation $q=\{q\}[q]$ for a quantity with units distinguishes
between the unit $[q]$ and the numerical value $\{q\}$.  For example,
for the speed of light, we have $c = 299\,792\,458$ m/s, where
$\{c\}=299\,792\,458$ and $[c]=$ m/s.  Evidently both of these factors
depend on the system of units, but the product $\{q\}[q]$ describes the
same physical quantity.  The factors $\{q\}$ and $[q]$ separately follow
the algebraic rules of multiplication and division, which allows for a
consistent dimensional analysis and conversion between different units.

In terms of this notation, the calculation in Eq.~(\ref{eq:simex}) can
be written as
\begin{eqnarray}
\{E\}[E] = \frac{1}{2}\,\{m\}[m]\left(\{v\}[v]\right)^2
\end{eqnarray}
or
\begin{eqnarray}
\{E\}\mbox{ J} 
= \frac{1}{2}\,\{m\}\{v\}^2
\mbox{ kg}\mbox{ m}^2\mbox{ s}^{-2}
= \frac{1}{2}\,\{m\}\{v\}^2 \mbox{ J} \, .
\end{eqnarray}
In this way, calculations are separated into a purely numerical part and
one involving units.  For nontrivial equations, working separately with
only the numerical values provides a practical way of carrying out the
calculation.  In particular, when mathematical functions such as
exponential, trigonometric, or Bessel functions are involved, the
arguments are necessarily numbers without units, and calculations are
done with only the numerical values.  This is further simplified if a
coherent system of units is used, so there is no additional numerical
factor.

Physical science is based on mathematical equations, which follow the
rules of analysis spelled out in numerous mathematical reference works.
Generally, in mathematical reference texts, distances, areas, and
angles, for example, are all dimensionless.  On the other hand, in
physical science, one uses units.

One consequence of this difference is that mathematics provides no
information on how to incorporate units into the analysis of physical
phenomena.  One role of the SI is to provide a systematic framework for
including units in equations that describe physical phenomena.

\section{Angles}
\label{sec:a}

Angles play an essential role in mathematics, physics, and engineering.
They fall into the category of quantities with dimensionless units in
the current SI, which leads to ambiguities in applications.  This issue
has been widely discussed in the literature, and arguments are given on
both sides of the question of whether angles are quantities that should
have units.\cite{1982033}

In part because units are rarely considered in mathematics, the unit of
radian for angles is rarely mentioned in the mathematics reference
literature, just as the meter is also rarely mentioned.  Units are
unnecessary in purely mathematical analysis.  By the same token, caution
is necessary in drawing conclusions about units based on purely
mathematical considerations.  For example, in the current SI, it is
stated that angles are dimensionless based on the definition that an
angle in radians is arc length divided by radius, so the unit is
surmised to be a derived unit of one, or a dimensionless unit.  However,
this reasoning is not valid, as indicated by the following example.  An
angle can also be defined as ``twice the area of the sector which the
angle cuts off from a unit circle whose centre is at the vertex of the
angle.'' \cite{1927002a}  This gives the same result for the numerical
value of the angle as the definition quoted in the SI Brochure, however
by following similar reasoning, it suggests that angles have the
dimension of length squared rather than being dimensionless.  This
illustrates that conclusions about the dimensions of quantities based on
such reasoning are clearly nonsense.  

Regardless of whether we view angles as having dimension or not, they
can be measured and the results can be expressed, for example, in units
of degrees, radians, or revolutions.  In elementary plane geometry or
daily life, degrees are usually used, and it is intuitively familiar to
think of a $45^\circ$ or $90^\circ$ angle and the fact that $360^\circ$
is a complete revolution.  In this case, the unit is degrees and
$[90^\circ] = \, ^\circ$.  

In calculus and physics it is convenient to use radians or rad for angle
units.  The angle in radians between two lines that cross at a point is
the length of circular arc $s$ swept out between the lines by a radius
vector of length $r$ from the crossing point divided by the length of
the radius vector.  The angle $\theta$ is thus given by
\begin{eqnarray}
\theta &=& \frac{s}{r}\mbox{ rad} \, ,
\label{eq:angsi}
\end{eqnarray}
which corresponds to $\{\theta\} = s/r$ and $[\,\theta] = \mbox{rad}$.

The conversion between radians and degrees follows from the relation
$360^\circ = 2\pi$ rad, which gives, for example,
\begin{eqnarray}
90^\circ = \frac{2\pi\mbox{ rad}}{360^\circ}\,90^\circ
= \frac{\pi}{2} \mbox{ rad},
\label{eq:conv}
\end{eqnarray}
where the rules of algebra are applied to the units to cancel degrees
from the equation.

In this context, units for angles obviously play a useful role.  As with
any measurable quantity, a given angle will have different numerical
values depending on the units in which the angle is expressed.  Units
such as degrees or radians are converted to other units by algebraic
calculations as in Eq.~(\ref{eq:conv}).  A problem with not including
any units for angles is illustrated by the equation corresponding to one
complete revolution
\begin{eqnarray}
2\pi \mbox{ rad} = 1 \mbox{ rev} \, ,
\label{eq:radrev}
\end{eqnarray}
where rev is the symbol for revolution.  If rad and rev were both
replaced by ``one'', as allowed in the current SI, then
Eq.~(\ref{eq:radrev}) would be nonsense.

In view of the problems that can occur if units for angles are omitted,
we consider the consequences of a consistent treatment of units for
angles in the following.  For an infinitesimal segment of a plane curve,
the change in angle ${\rm d} \theta$ of the tangent to the curve is
proportional to the infinitesimal change in position ${\rm d} s$ along
the curve, where we define the constant of proportionality to be the
angular curvature ${\cal C}$:
\begin{eqnarray}
{\rm d}\theta &=& {{\cal C}}\, {\rm d}s \, ,
\end{eqnarray}
where ${\cal C}$ has units of rad/m.  Evidently, the angular curvature is a
measure of the amount of bending of the segment of the curve.  (This is
different from curvature of a graph in elementary calculus or the
curvature in differential geometry both of which are dimensionless and
do not involve angles.)  If an angular radius of curvature ${\cal R}$ is
defined as
\begin{eqnarray}
{\cal R} &=& \frac{1}{{\cal C}}
\label{eq:roc}
\end{eqnarray}
then
\begin{eqnarray}
{\rm d}\theta &=& \frac{{\rm d}s}{{\cal R}} \, .
\end{eqnarray}
The quantity ${\cal R}$ with units m/rad should be distinguished from $r$ in
Eq.~(\ref{eq:angsi}) which has units of m.  If the curve is a portion of
a circle, then we have
\begin{eqnarray}
\theta &=& \frac{s}{{\cal R}} \, ,
\label{eq:angnew}
\end{eqnarray}
in analogy with Eq.~(\ref{eq:angsi})  and $\{{\cal R}\} = \{r\}$.  In fact,
Eq.~(\ref{eq:angsi}) is the same as Eq.~(\ref{eq:angnew}) if the
replacement rad $\rightarrow 1$ is made.  

This extends naturally to steradians for solid angle, abbreviated sr,
for which an infinitesimal solid angle subtended by the area $\rd a$ on
the surface of a sphere is given by
\begin{eqnarray}
\rd \iO = \fr{\rd a}{{\cal R}^2} \, ,
\end{eqnarray}
which has units rad$^2$.  Table~\ref{tab:aq} compiles a number of quantities
involving angles and the associated units.

\begin{table}[h]
\caption{Quantities involving angles and their units. \label{tab:aq}}
\begin{tabular}{@{\quad}l@{\hspace{50pt}}l@{\hspace{50pt}}l@{\quad}}
\toprule
Quantity \qquad & Equation \qquad & Units \\[3 pt]
\colrule
angle & $\theta$ & rad \\[3 pt]

angular curvature & ${\cal C} = \fr{\rd \theta}{\rd s} $ & rad m$^{-1}$
\\[3 pt]

angular radius of curvature & ${\cal R} = \fr{1}{{\cal C}} = \fr{\rd s}{\rd
\theta}$ & m rad$^{-1}$ \\[3 pt]

infinitesimal arc length  & $\rd s = {\cal R} \, \rd \theta$ & m \\[3 pt]

infinitesimal angle  & $\rd \theta = \fr{\rd s}{{\cal R}}$ & rad \\[3 pt]

solid angle & $\iO$  & sr \\[3 pt]

infinitesimal surface area & $\rd a$  & m$^2$ \\[3 pt]

infinitesimal solid angle  &  $\rd \iO = \fr{\rd a}{{\cal R}^2} $ & sr
=
rad$^2$ \\[3 pt]

\botrule
\end{tabular}
\end{table}

In applications, angles appear in the exponential and trigonometric
functions, and these functions are defined for an argument that is a
dimensionless number, {\it i.e.,} the numerical value of the angle
expressed in radians.  The exponential function is given by its power
series
\begin{eqnarray}
\re^x &=& 1 + x + \frac{x^2}{2} + \dots ,
\label{eq:exp}
\end{eqnarray}
and the relation 
\begin{eqnarray}
\re^{\ri y} = \cos{y} + \ri \sin{y}
\label{eq:compn}
\end{eqnarray}
follows from the series expansions of the cosine and sine functions.
The unit ``radian'' cannot be included as a factor in the arguments of
these functions, because every term in the power series must have the
same unit.

The connection of these functions to angles follows from the fact that
Eq.~(\ref{eq:compn}) is a point in the complex plane on the unit circle
at an angle $\theta = y$ rad in the counter-clockwise direction from the
positive real axis.  The periodicity of the function $\re^{\ri y}$ fixes
the unit of the angle to be $[\theta] = $ rad, because both the angle
$\theta = y$ rad and the function $\re^{\ri y}$ go through one complete
cycle as $y = \{\theta\}$ ranges from 0 to $2\pi$.  The choice of any
other unit for $[\theta]$ would not align these two periods.  

However, it is the general practice in physics to write the exponential
function of an angle $\theta = y$ rad as $\re^{\ri \theta}$ rather than
$\re^{\ri y}$ or $\re^{\ri \{\theta\}}$.  In fact, in carrying out
calculations, scientists do not usually distinguish between $\theta$ and
$\{\theta\}$, which amounts to treating rad as being 1.

This reveals a conflict between consistent application of dimensional
analysis and common usage.  A consistent application of dimensional
analysis is needed in order for the SI to be used as the basis for any
computer algebra program that takes units into account.
\cite{1997199,2001383} This is likely to be an increasingly popular way
of doing calculations, and having a consistent foundation is necessary
to prevent errors.  For such applications, it is important to use the
numerical value of angles when expressed in radians, $\theta$/rad,
in exponential and trigonometric functions as well as more general
functions of angles in mathematical physics to avoid errors of $2\pi$
which might otherwise occur.  (For example, one could confuse Hz and
rad/s as described in Sec.~\ref{sec:pp}.)  On the other hand, for
general use in printed equations following the common practice, the
argument of the exponential and trigonometric functions is simply
written as $\theta$ which corresponds to replacing rad by 1.  Of course,
this replacement can only be done for the unit rad, and not revolutions
or degrees, replacements that would introduce numerical factors.  In
this sense, the unit rad is a coherent unit in the SI, whereas
revolutions and degrees are not.

\section{Periodic phenomena}
\label{sec:pp}

Periodic phenomena in physics include rotations of an object, cycles or
repetitions of a wave, or a series of any regular, repetitive events.
Such periodic phenomena are characterized by a frequency whose units can
be an angular factor or a cycle divided by time.  In the SI,
cycles/second = cyl/s is named hertz or Hz, and
\begin{eqnarray}
1 \mbox{ Hz} = 1 \, \mbox{cyl}\mbox{ s}^{-1}
= 2 \pi \, \mbox{rad}\mbox{ s}^{-1} \, ,
\label{eq:hzrad}
\end{eqnarray}
where the second equality follows from Eq.~(\ref{eq:radrev}) with
revolution replaced by cycle.  Hz may be viewed as being equivalent to
rotations per second, but often, ``rotations'' is used for mechanical
motion and ``cycles'' is used for waves.  

The symbol used for angular frequency is $\omega$, which is understood
to mean the frequency in units of rad/s, while the symbols $\nu$ or $f$
are used to denote frequency expressed in hertz.  The relation between
the numerical value of a particular frequency expressed in Hz or rad
s$^{-1}$ is given by
\begin{eqnarray}
\{\nu\}_{\rm Hz}[\mbox{Hz}] = \{\omega\}_{\rm rad~
s^{-1}}[{\rm rad\cdot s}^{-1}]
=\frac{\{\omega\}_{\rm rad\cdot s^{-1}}}{2\pi}\,[\mbox{Hz}] \, ,
\label{eq:fhzwrad}
\end{eqnarray}
or
\begin{eqnarray}
\frac{\{\omega\}_{\rm rad\cdot s^{-1}}}{2\pi} =  \{\nu\}_{\rm Hz} \, ,
\label{eq:radhz}
\end{eqnarray}
where the second equality in Eq.~(\ref{eq:fhzwrad}) follows from
Eq.~(\ref{eq:hzrad}).  As already noted, radians behave as coherent
units for the SI, so we make the identification $\{\omega\}_{\rm
rad\cdot s^{-1}}=\{\omega\}$, where the curly brackets with no subscript
indicate that the numerical value corresponds to a coherent SI
unit.  However, a consequence of this convention is that the unit Hz is
not a coherent SI unit as indicated by Eq.~(\ref{eq:radhz}).  This is in
conflict with the current SI where Hz is treated as a coherent SI unit,
only because cyl is replaced by ``one''.  Since this leads to an
inconsistency, we propose that the SI be modified in such a way that Hz
is neither treated as a coherent SI unit nor replaced by s$^{-1}$.

We note that if both rad and cyl are replaced by ``one'', as allowed in the
current SI, then Eq.~(\ref{eq:radhz}) takes the form of the
(questionable) relation
\begin{eqnarray}
\omega \ \substack{?\\=\\ \vbox to 2 pt{}}\ 2\pi \nu \, ;
\label{eq:wsimnu}
\end{eqnarray}
we employ the symbol $\substack{?\\=\\ \vbox to 1 pt{}}$ to emphasize
that the equation is only true with those inappropriate replacements.
In fact, the correct equation is given by Eq.~(\ref{eq:radhz}) which
only involves the numerical values.  We recognize that when people write
Eq.~(\ref{eq:wsimnu}) as an equality, they mean what is stated in
Eq.~(\ref{eq:radhz}).  In other words, when for a given frequency people
mistakenly write $\omega$ is equal to $2\pi\nu$, they correctly mean
that the numerical value in radians per second of $\omega$ is
$2\pi$ times the numerical value in hertz of $\nu$.

A basic equation for waves is the relation between the wavelength
and the frequency.  This is generally written
\begin{eqnarray}
\lambda\nu = c \, ,
\label{eq:ln}
\end{eqnarray}
where $\lambda$ is the crest to crest wavelength and $c$ is the wave
velocity, which for electromagnetic radiation in free space is the speed
of light.  From the requirement that units on both sides of an equality
must be the same, and the conventions that $c$ has the unit m/s in the
SI and $\nu$ has the unit Hz, Eq.~(\ref{eq:ln}) implies that the unit
for $\lambda$ is
\begin{eqnarray}
[\lambda] = \frac{[c]}{[\nu]} = \mbox{m s}^{-1}\mbox{Hz}^{-1}
= \mbox{m cyl}^{-1} \, ,
\label{eq:lamu}
\end{eqnarray}
which has a self-evident intuitive interpretation. Neither Hz nor m/cyl
is a coherent unit.  For a ``coherent'' version of Eq.~(\ref{eq:ln}),
that is, an equation in which $c$ has the unit m/s and the frequency has
the unit rad/s, we write
\begin{eqnarray}
\lambdabar\omega = c
\label{eq:lbo}
\end{eqnarray}
which implies that the reduced wavelength $\lambdabar$ has the units
\begin{eqnarray}
[\lambdabar] = \frac{[c]}{[\omega]} = \frac{\rm m~s^{-1}}
{{\rm rad~s^{-1}}} = \mbox{m rad}^{-1}
\label{eq:lb}
\end{eqnarray}
and that
\begin{eqnarray}
\{\lambdabar\} = \frac{\{\lambda\}_{\rm m\cdot cyl^{-1}}}{2\pi} \, ,
\label{eq:lbar}
\end{eqnarray}
where as before, the absence of a subscript on the curly brackets
indicates that the numerical value refers to SI units, which in
this case are m rad$^{-1}$.  Again, when the relation
\begin{eqnarray}
\lambdabar  \ \substack{?\\=\\ \vbox to 2 pt{}}\ \frac{\lambda}{2\pi} 
\end{eqnarray}
is treated as an equality, what is meant is Eq.~(\ref{eq:lbar}).

Another quantity associated with waves is the wave vector
\begin{eqnarray}
k = \frac{1}{\lambdabar} \, .
\end{eqnarray}
It has units of radians per meter or rad/m.  (The magnitude of the wave
vector is to be distinguished from the wavenumber $\lambda^{-1}$ used in
spectroscopy.) With these units for $k$, the covariant phase $kx-\omega
t$ for a wave propagating in the $x$ direction, where $x$ is a
coordinate and $t$ is the time, is homogeneous in the unit rad.  This is
consistent with the quantum mechanical expression for momentum.

In classical mechanics, rotational motion of a rigid body can be
described by an angle $\theta$ about a fixed axis of rotation as a
function of time $t$, an angular velocity $\omega$
\begin{eqnarray}
\omega = \frac{\rd \theta}{\rd t} \, ,
\end{eqnarray}
and an angular acceleration $\alpha$, given by 
\begin{eqnarray}
\alpha = \frac{\rd \omega}{\rd t} = \frac{\rd^2 \theta}{\rd t^2} \, .
\end{eqnarray}
Evidently, $\theta$, $\omega$, and $\alpha$ have units of rad, rad
s$^{-1}$, and rad s$^{-2}$.  Units for other quantities associated with
rotational motion, such as the moment of inertia, may be deduced from
the defining equations.  As a rule of thumb, in order to obtain a
coherent set of units it is necessary to take the radius $r$ that
appears in such expressions to be the angular radius of curvature ${\cal R}$
with units m/rad defined in Eq.~(\ref{eq:roc}).  Table~\ref{tab:mq}
lists various quantities associated with rotational motion of a point
mass at a distance $r$ from the axis of rotation, the relevant
equations, and the corresponding units.  A longer list is given by
Eder.\cite{1982033}

\begin{table}[h]
\caption{Quantities involving rotational motion and their units. 
\label{tab:mq}}
\begin{tabular}{@{\quad}l@{\hspace{50pt}}l@{\hspace{50pt}}l@{\quad}}
\toprule
Quantity \qquad & Equation \qquad & Units \\[3 pt]
\colrule

angular velocity & $\omega = \fr{\rd \theta}{\rd t} $ & rad
s$^{-1}$ \\[3 pt]

angular acceleration & $\alpha = \fr{\rd \omega}{\rd t} = 
\fr{\rd^2 \theta}{\rd t^2} $ & rad s$^{-2}$ \\[3 pt]

velocity & $v = \fr{\rd s}{\rd t} = {\cal R} \,\fr{\rd \theta}{\rd t} 
= {\cal R} \, \omega$ & m s$^{-1}$ \\[3 pt]

moment of inertia & $I = m {\cal R}^2 $ & kg m$^2$ rad$^{-2}$
\\[3 pt]

angular momentum & $L = I \omega = m {\cal R}^2 \omega$ & J s rad$^{-1}$
\\[3 pt]

torque & $N = I \alpha$ & J rad$^{-1}$
\\[3 pt]

energy & $E =  I\,\frac{\textstyle\omega^2}{\textstyle 2}$  & J \\[3 pt]

centrifugal force & $ F_{\rm c} = m\omega^2{\cal R} = \fr{mv^2}{{\cal R}}$ & N
rad \\[3 pt]

\botrule
\end{tabular}
\end{table}

In electromagnetism and quantum mechanics, the product $\omega t$ of
angular frequency and time often appears in the exponential function.
This is similar to the case for angles as discussed in Sec.~\ref{sec:a}.
In quantum mechanics for example, it is conventional to write
\begin{eqnarray}
\re^{-\ri\,\omega t} \, ,
\end{eqnarray}
where actually what is meant is
\begin{eqnarray}
\re^{-\ri\{\omega t\}} \, .
\end{eqnarray}
Here, as for angles, it is common practice to treat rad as ``one'',
which does not lead to problems because rad is a coherent SI unit.

\section{Counting quantities}
\label{sec:cq}

Many scientific applications involve counting of events or entities.
For example, in radioactive decay, events occur at random times, but
still have a well-defined rate when averaged over a sufficiently long
time with a large enough sample.  The result of a measurement, where
decays trigger counts in a detector, is counts/second or cnt/s.  The SI
unit for activity of a radiative sample is becquerel or Bq, meaning
decays per second, which is related to counts per second through the
overall detection efficiency.  However, in the current SI, it is said
that the becquerel has units of s$^{-1}$, which means that the decay or
count in the numerator is dropped.  Here we take issue with this
prescription and argue that the unit ``decay'' or ``count'' should be
retained, because it provides information about the number that precedes
it in the expression for the quantity.  In addition, since the current
SI replaces both Hz and Bq by s$^{-1}$, the distinction between these
units is lost and sometimes leads to the dangerous and sadly mistaken
use of Hz, which refers to periodic cycles, for the rate of random
events.  (Non-radioactive decay {\it e.g.,} decay of excited atomic
states, is similarly a random process and is properly measured in decays
per second, but not traditionally in Bq, and certainly not in Hz.)

This is a special case of counting in general.  Things that can be
counted include events, such as decays or clicks of a detector, and
entities, such as atoms or molecules.  For such countable things, it is
useful to include a designation of what is being counted in the unit for
the corresponding quantities.  Quantities involving counting are not
restricted to numbers and rates.  For example, if in a certain time
interval there are ${\cal D}=200$ decays $=200$ dcy and the detector
registers ${\cal N}=20$ counts $=20$ cnt, then the efficiency $\eta$ of
the detection is
\begin{eqnarray}
\eta = \frac{{\cal N}}{{\cal D}} = 0.1 \, 
\frac{\mbox{cnt}}{\mbox{dcy}} \, .
\end{eqnarray}
Conversion between the count rate and the decay rate may be made using
the detection efficiency as a conversion factor.  For this detector, if
a count rate of ${\cal Q} = 73$ cnt/s is observed, it indicates a decay
rate ${{\it \Gamma}}$ given by
\begin{eqnarray}
{\it \Gamma} = \frac{\cal Q}{\eta} = \frac{73 \mbox{ cnt/s}}{0.1 \mbox{
cnt/dcy}} = 730 \, \frac{\mbox{dcy}}{\mbox{s}} \, .
\end{eqnarray}
In this case, the detector efficiency has units, unlike the
recommendation of the current SI where it would be simply a number.  The
units provide useful information in a form that can be incorporated into
calculations.

Counting also applies to entities such as atoms or molecules.  The
number density $n$ of molecules in a given volume is the number of
molecules ${\cal M}$ divided by the volume $V$
\begin{eqnarray}
n = \frac{{\cal M}}{V} \, ,
\label{eq:numd}
\end{eqnarray}
which in the current SI has units of m$^{-3}$.  However, this is another
case where specification of what the density refers to is useful.  This
would make the number density consistent with other forms of density,
such as mass density or charge density, which have units of kg/m$^3$ and
C/m$^3$, respectively.  For number density, the unit should be mcl/m$^3$,
which follows naturally when ${\cal M}$ has the unit mcl, where mcl
is the unit for the number of molecules.  For macroscopic numbers of
molecules or atoms, it is convenient to use the unit mole or mol, where
\begin{eqnarray}
1 \mbox{ mol} = 6.02\dots\times10^{23} \mbox{ ent} \, ,
\label{eq:nd}
\end{eqnarray}
where ent is the suggested symbol for entity.  This expression makes it
clear that the mole, which is the unit of amount of substance, is not
just a number, but a number of entities.  This relation can be used as a
conversion factor between number density and molar density, which differ
only in their units.  As an example
\begin{eqnarray}
n =
2.5\times10^{25} \,\frac{\mbox{mcl}}{\mbox{m}^3} = \frac{1\mbox{ mol}}
{6.02\dots\times10^{23}\mbox{ mcl}}
\, 2.5\times10^{25}\,\frac{\mbox{mcl}}{\mbox{m}^3}
= 42\,\frac{\mbox{mol}}{\mbox{m}^3} \, .
\label{eq:md}
\end{eqnarray}
The presence of units makes the conversion more clear than it would be
if the unit mcl were absent from Eq.~(\ref{eq:md}) as
the current SI prescribes, and for polyatomic molecules removes any
ambiguity about whether atoms or molecules are being counted.

A list of suggested unit names for events and entities is given in
Table~\ref{tab:cq}.  Other items can be named as needed.

\begin{table}[h]
\caption{Quantities involving counting and their unit symbols. 
\label{tab:cq}}
\begin{tabular}{@{\quad}l@{\hspace{80pt}}l@{\quad}}
\toprule
Quantity \qquad & Unit symbol \\[3 pt]
\colrule
events & evt \\[3 pt]

\quad number of counts & \quad cnt \\[3 pt]

\quad number of decays & \quad dcy \\[3 pt]

entities & ent \\[3 pt]

\quad number of molecules & \quad  mcl \\[3 pt]

\quad number of atoms & \quad  atm \\[3 pt]

\quad number of particles & \quad  pcl \\[3 pt]

\botrule
\end{tabular}
\end{table}

\section{Fundamental constants}
\label{sec:fc}

Fundamental constants are parameters in the equations that describe
physical phenomena and have the units that are necessary for dimensional
consistency.  The CODATA recommended values and units for the
constants\cite{2012158} are based on the conventions of the current SI,
and any modifications of those conventions will have consequences for
the units.  

For example, the equation
\begin{eqnarray}
E = \hbar \omega 
\end{eqnarray}
relates $E$, the energy of a photon, with its angular frequency
$\omega$.  These quantities are related through the Planck constant
$\hbar$, and for the equation to be dimensionally consistent, taking
into account the modifications of the SI under consideration, the unit
of $\hbar$ must be J s rad$^{-1}$, or more suggestively, J/(rad
s$^{-1}$).  This is in contrast with the CODATA tabulated value for
$\hbar$ which has the unit J s.  Similarly, the equation
\begin{eqnarray}
E = h \nu \, ,
\end{eqnarray}
where $\nu$ is the photon frequency in hertz, implies that the unit for $h$
is J Hz$^{-1}$.  Both J s rad$^{-1}$ and J Hz$^{-1}$ reduce to J s in
the current SI, but they are distinct when units are treated
consistently.  The two expressions for the photon energy for a given
frequency imply
\begin{eqnarray}
\hbar\omega = h\nu \, ,
\end{eqnarray}
and together with Eq.~(\ref{eq:radhz}) lead to the conventional relation
\begin{eqnarray}
\{\hbar\} = \frac{\{h\}_{\rm J\cdot Hz^{-1}}}{2\pi} \, ,
\label{eq:hbarco}
\end{eqnarray}
between the numerical values of the Planck constant expressed in
different units.  One often sees
\begin{eqnarray}
\hbar  \ \substack{?\\=\\ \vbox to 2 pt{}}\  \frac{h}{2\pi} \, ,
\label{eq:hbarq}
\end{eqnarray}
but as before, when Eq.~(\ref{eq:hbarq}) is treated as an equality, what
is meant is Eq.~(\ref{eq:hbarco}).

Another basic constant involving $\hbar$ is the reduced Compton
wavelength of the electron $\lambdabar_{\rm C}$ given by
\begin{eqnarray}
\lambdabar_{\rm C} = \frac{\hbar}{m_{\rm e}c}
\end{eqnarray}
which has the units
\begin{eqnarray}
[\lambdabar_{\rm C}] = \frac{[\hbar]}{[m_{\rm e}][c]} = 
\mbox{m rad}^{-1}
\end{eqnarray}
consistent with Eq.~(\ref{eq:lb}).  Similarly, the Bohr radius $a_0$ is
related to the reduced Compton wavelength by
\begin{eqnarray}
\lambdabar_{\rm C} = \alpha a_0 \, ,
\end{eqnarray}
where $\alpha$ is the dimensionless fine-structure constant, so that 
\begin{eqnarray}
[a_0] = \mbox{m rad}^{-1} \, ,
\end{eqnarray}
which is consistent with the use of the angular radius of curvature for
mechanical rotational motion.  For the Rydberg constant, the definition
\begin{eqnarray}
R_\infty = \frac{\alpha}{4\pi a_0}
\end{eqnarray}
suggests the units
\begin{eqnarray}
[R_\infty] = \mbox{cyl m}^{-1}
\end{eqnarray}
in order to be consistent with the Rydberg formula
\begin{eqnarray}
\frac{1}{\lambda} = R_\infty\left(\frac{1}{n_1^2} - \frac{1}{n_2^2}\right)
\, .
\end{eqnarray}
A corresponding angular version of the Rydberg constant is given by
\begin{eqnarray}
 \raisebox{6 pt}{\line(1,0){10}}\mskip-16mu R_\infty 
= \frac{\alpha}{2 a_0} \, ,
\end{eqnarray}
with units rad m$^{-1}$, where
\begin{eqnarray}
\{\raisebox{6 pt}{\line(1,0){10}}\mskip-16mu R_\infty\}
= 2\pi \{R_\infty\}\, .
\end{eqnarray}

The expression for the fine-structure constant $\alpha$ in the current
SI is given by
\begin{eqnarray}
\alpha = \frac{e^2}{4\pi\epsilon_0\hbar c} \, ,
\label{eq:alpha}
\end{eqnarray}
where $e$ is the unit charge and $\epsilon_0$ is the vacuum permittivity
(electric constant).  The $\hbar$ in that expression may be seen to
arise from the form of electromagnetic interactions in the Schr\"odinger
equation as follows.  For the hydrogen atom
\begin{eqnarray}
\left[\frac{\bm p^2}{2m_{\rm e}}+V(\bm x)\right]\psi(\bm x) 
= E\psi(\bm x) \, ,
\end{eqnarray}
where $\bm p = -\ri\hbar\bm\nabla$, $V(\bm x) = -e^2/(4\pi\epsilon_0|\bm
x|)$, $\psi(\bm x)$ is the wave function, and $E$ is the energy
eigenvalue.  If the coordinate is written as a dimensionless factor
times the reduced Compton wavelength of the electron $\bm x = \tilde {\bm
x}
\,\hbar/(m_{\rm e}c)$, then the equation is of the completely
dimensionless form
\begin{eqnarray}
\left[-\frac{\tilde{\bm \nabla}^2}{2} - \frac{\alpha}{|\tilde{\bm
x}|}\right]\tilde\psi(\tilde{\bm x}) = \tilde E \, \tilde\psi(\tilde{\bm
x}) \, ,
\end{eqnarray}
where $\tilde E = E/m_{\rm e}c^2$ and $\tilde\psi(\tilde{\bm x}) = \psi(\bm
x\,m_{\rm e}c/\hbar)$.  However when an inverse radian is included in
$\hbar$, Eq.~(\ref{eq:alpha}) must be modified in order for $\alpha$ to
be dimensionless.  This can be done, although not uniquely, by using the
freedom in the definition of electrical quantities as discussed by
Jackson\cite{1998165} in his appendix on units and dimensions.  If
replacements to the definitions of the unit factors given by $k_1
\rightarrow k_1/\mbox{rad}$ and $k_2 \rightarrow k_2/\mbox{rad}$ are
made, then there is no change to the SI form of the Maxwell equations
other than the modification of the units of $\epsilon_0$ and $\mu_0$ to
be
\begin{eqnarray}
\left[\epsilon_0\right] = \left[\frac{e^2}{\hbar c}\right] = {\rm
C^2~J^{-1}~rad~m^{-1}}  \, .
\end{eqnarray}
and
\begin{eqnarray}
\left[\mu_0\right] = \left[\frac{\hbar}{e^2c}\right]
= {\rm kg~m~rad^{-1}~C^{-2}}
\end{eqnarray}
where $\mu_0$ is the vacuum permeability (magnetic constant).

We now turn to constants related to counting.  The Avogadro constant
$N_{\rm A}$ is the number of entities in one mole which can be written
as
\begin{eqnarray}
N_{\rm A} = 6.02\dots\times10^{23} \mbox{ ent mol}^{-1} \, ,
\end{eqnarray}
in accord with Eq.~(\ref{eq:nd}).  Evidently, this constant can be
viewed as the conversion factor between entities and moles.  It also
provides the relation between the molar gas constant $R = 8.31\dots$ J
mol$^{-1}$ K$^{-1}$ and the Boltzmann constant $k$, which is thus given
by
\begin{eqnarray}
k = \frac{R}{N_{\rm A}} = 1.38\dots\times10^{-23} 
{\rm ~J~K^{-1}~ent^{-1}} \, .
\end{eqnarray}
Similarly, the Avogadro constant relates the Faraday constant $F =
9.64\dots\times10^{4}$ C mol$^{-1}$ to the unit charge, which can be
written as
\begin{eqnarray}
e = \frac{F}{N_{\rm A}} = 1.60\dots\times10^{-19} {\rm ~C~ent^{-1}} \, .
\end{eqnarray}
Evidently, this expression allows for an explicit conversion between
number density and charge density, which takes the form $[\rho] = [{\rm
C~m^{-3}}] = [e\,n]$, with $n$ as defined in Eq.~(\ref{eq:numd}).

This is not an exhaustive list of fundamental constants that are
affected by the explicit expression of radians or entities, but the
pattern for including such units should be clear.  On the other hand,
the majority of constants remain unchanged.

\begin{table}[h]
\caption{Fundamental constants and their units. \label{tab:fc}}
\begin{tabular}{@{\quad}l@{\hspace{50pt}}l@{\hspace{50pt}}l@{\quad}}
\toprule
Constants \qquad & Symbol \qquad & Units \\[3 pt]
\colrule
Reduced Planck constant & $\hbar$ & J s rad$^{-1}$ \\[3 pt]

Planck constant & $h$ & J Hz$^{-1}$ \\[3 pt]

Electron reduced Compton wavelength & $\lambdabar_{\rm C}$ 
   & m rad$^{-1}$ \\[3 pt]

Electron Bohr radius & $a_0$ & m rad$^{-1}$ \\[3 pt]

Rydberg constant & $
 \raisebox{5 pt}{\line(1,0){10}}\mskip-15mu R_\infty 
$ & rad m$^{-1}$ \\[3 pt]

vacuum permittivity (electric constant) & $\epsilon_0$ & 
${\rm C^2~J^{-1}~rad~m^{-1}}$ \\[3 pt]

vacuum permeability (magnetic constant) & $\mu_0$ & 
${\rm kg~m~rad^{-1}~C^{-2}}$ \\[3 pt]

Avogadro constant & $N_{\rm A}$ & ent mol$^{-1}$ \\[3 pt]

Boltzmann constant & $k$ & ${\rm J~K^{-1}~ent^{-1}}$ \\[3 pt] 

elementary charge & $e$ & ${\rm C~ent^{-1}}$ \\[3 pt]

\botrule
\end{tabular}
\end{table}

\section{Conclusion and Recommendations}
\label{sec:cr}

Modifications of the SI to eliminate the incoherence that results from
dropping so-called dimensionless quantities have been identified and
discussed.  There is some latitude in how the modifications might be
taken into account by users of the SI.  However, one conclusion that is
not optional is that the unit hertz cannot be regarded as a coherent
unit of the SI, in contrast to its designation in the current form of
the SI, where cycles are ignored and Hz may be replaced by s$^{-1}$.

At the same time, we have shown that the unit radian can play a useful
role in providing consistency of units and should be regarded as the
coherent unit for angles in the SI.  Therefore, we recommend that
quantities involving rotation, angles, or angular frequencies be
reported including radians as a unit.  However, we do not recommend that
one change the common practice of writing expressions like $\cos{\omega
t}$ to the more pedantic form of $\cos{\{\omega t\}}$.  It would be too
disruptive to make it a requirement of the SI to distinguish between an
angle and its numerical value.

For units involved in counting, the prevailing practice is to include
them in expressions for such quantities.  This is in contrast to the
current SI, where they are omitted.  Here a consistent formulation for
the use of such quantities is provided.

With regard to fundamental constants, publications of the CODATA Task
Group on Fundamental Constants are based on the current
SI.\cite{2012158}  We recommend that future listings of values of the
fundamental constants give complete units, including radians and
counting units in order to provide a guide for a consistent use of the
constants, particularly by computer programs that include units.  Users
of the constants may still choose to omit either radians or counting
units, but including them in the listed values would encourage users to
use them coherently if they choose to.

\section{Acknowledgement}

We are grateful to Prof. Ian Mills for many valuable conversations
and for contributing seminal ideas related to this paper.

\end{document}